\documentclass[showpacs, preprintnumbers, nofootinbib, aps, prl, superscriptaddress,10pt, showkeys, notitlepage, 
aastex]{revtex4-2}

\usepackage{graphicx,amssymb,amsmath,amsthm,amsfonts,epsfig, setspace}

\usepackage[linktocpage]{hyperref}
\usepackage[usenames,dvipsnames]{color}
\usepackage{epstopdf}
\usepackage{pifont}
\definecolor{darkred}{rgb}{0.5,0,0}
\definecolor{darkgreen}{rgb}{0,0.5,0}
\definecolor{darkblue}{rgb}{0,0,0.5}
\definecolor{prussian}{rgb}{0.0, 0.19, 0.33}
\definecolor{richelectricblue}{rgb}{0.03, 0.57, 0.82}
\definecolor{teal}{rgb}{0.0, 0.5, 0.5}
\definecolor{mediumseagreen}{rgb}{0.24, 0.7, 0.44}
\definecolor{lust}{rgb}{0.9, 0.13, 0.13}
\definecolor{ballblue}{rgb}{0.13, 0.67, 0.8}
\definecolor{darkcyan}{rgb}{0.0, 0.55, 0.55}
\definecolor{mountainmeadow}{rgb}{0.19, 0.73, 0.56}
\definecolor{palecarmine}{rgb}{0.69, 0.25, 0.21}
\definecolor{richcarmine}{rgb}{0.84, 0.0, 0.25}
\definecolor{tangelo}{rgb}{0.98, 0.3, 0.0}
\definecolor{venetian}{rgb}{0.784,0.031,0.082}
\definecolor{bdfrance}{rgb}{0.192,0.549,0.906}

\hypersetup{colorlinks=true, citecolor=venetian,
linkcolor=bdfrance, urlcolor=lust}
\usepackage{amsmath,amssymb}
\usepackage{tensor}
\usepackage{mathtools}
\usepackage{amsbsy}
\usepackage{bm}
\usepackage{float}

\usepackage{cleveref}

\usepackage{appendix}

\newcommand{\be}{\begin{equation}}
\newcommand{\ee}{\end{equation}}
\newcommand{\bear}{\begin{eqnarray}}
\newcommand{\eear}{\end{eqnarray}}

\begin{document}

\title{Einstein ring fingerprint around supermassive black holes illuminated by broad line region spectral lines}

\author{Konstantinos Kostaros}
\email{kkostaro@auth.gr}
\affiliation{Department of Physics, Aristotle University of Thessaloniki, Thessaloniki 54124, Greece}

\author{George Pappas}
\email{gpappas@auth.gr}
\affiliation{Department of Physics, Aristotle University of Thessaloniki, Thessaloniki 54124, Greece}

\author{Padelis Papadopoulos}
\email{padelis@auth.gr}
\affiliation{Department of Physics, Aristotle University of Thessaloniki, Thessaloniki 54124, Greece}
\affiliation{Research Center for Astronomy, Academy of Athens, Soranou Efesiou 4, GR-11527 Athens, Greece}

\author{Wing-Fai Thi}
\affiliation{Karpfenstrasse 18, 81825 Munich, Germany}

\begin{abstract} The continuum emission from the hot and ionized inner regions of  accretion disks around supermassive black holes (SMBHs), which is strongly lensed by the light ring (i.e., the unstable photon orbit) is always superimposed on that from the locally emitting plasma near the innermost stable circular orbit,  masking strong-gravity effects and making their study difficult. Cleaner images of (strong-lensing)-generated brightness distributions set against a nonluminous background would thus be preferable. An SMBH illumination source that allows for this could be the more distant, cooler, neutral gas reservoir, underlying the so-called broad line region. Such a far-field illumination is ideal for producing the aforementioned images, and thus better suited for strong gravity tests. In such an illumination scenario, however, two prominent first-order effects stand out, which will be the focus of this Letter: an hourglass-like brightness distribution around the SMBH, and an associated double-peaked inner feature imprinted on the disk's global double-horned line profile, for lines of sight near the accretion disk plane. The latter effect can be discerned even in unresolved sources where only the total line profile is accessible. Interferometers operating in spectral line mode can reveal this hourglass image {\it and} the inner double peak spectral feature it imprints on the global line profile, and yield SMBH mass estimates for edge-on systems. 
\end{abstract} 
  
\maketitle

\emph{Context.  ---} 
%
%
The first images of the shadows of supermassive black holes (SMBHs) \cite{Akiyama_2019,EHT_M87_2,EHT_M87_3,EHT_M87_4,EHT_M87_5,EHT_M87_6,EHT_Sgr1,EHT_Sgr2,EHT_Sgr3,EHT_Sgr4,EHT_Sgr5,EHT_Sgr6,EHT_M87_7}, were produced using  Very Long Baseline Interferometry imaging by the Event Horizon Telescope collaboration (EHT), and opened a new avenue of exploration on black hole (BH) physics. The EHT images of the shadows of the BHs M87* and Sgr A* \cite{Akiyama_2019,EHT_Sgr1}, encode information for the structure of the spacetime near the light ring, i.e., the vicinity of the unstable photon orbit, but also of the astrophysical environment, i.e., the accretion region near the innermost stable circular orbit (ISCO) \cite{Volkel:2020xlc,Glampedakis:2021oie,Glampedakis:2023eek,Gralla_2020,Younsi:2021dxe,Bauer:2021atk,Glampedakis:2023eek,Gralla:2020nwp,Lima:2021las,Medeiros:2019cde,Gralla:2019xty,Johannsen_2010,Olmo:2023lil}. The future of these methods is promising, but they nevertheless contain some fundamental limitations. 

The shadow of an SMBH against the continuum emission of the hot inner accretion disk, which could be used for strong gravity tests, incorporates the interplay between the complex accretion processes taking place in those regions {\it and} the complicated strong gravity scattering of the radiation emitted by the disk towards the BH's unstable photon orbit. Moreover, the disk continuum emission from the immediate vicinity of the SMBH, emanating from matter distributed all the way down to the ISCO, will have some unwanted characteristics. Since the ISCO is close to the light-ring and defines a region where brightness drops relatively abruptly, direct emission from that region will inevitably contaminate the radiation emanating from the strong lensing region, burdening the extraction of information on strong-gravity physics with the complications of accretion disk physics \cite{Gralla:2019xty,Glampedakis:2021oie,Ozel:2021ayr,Younsi:2021dxe,Lara:2021zth,Urso:2025gos}. 
For example, tests of strong-gravity that rely on the precise determination of the size and shape of the BH shadow will suffer from the brightness of the accretion disk near the ISCO \cite{Volkel:2020xlc,Gralla_2020,Younsi:2021dxe,Fernandes:2024ztk,Gyulchev:2024iel,DeliyskiPhysRevD.111.064068,BenAchour:2025uzp}.  

Therefore, for precision strong-gravity tests using the shadows of BHs and their light rings \cite{Kostaros_2022,Kostaros:2024vbn}, one must circumvent these difficulties. A novel idea is to use a BH-illuminating source that lies farther away from the strong lensing areas and emits spectral line, rather than continuum emission, and without such line emission emanating from the hotter inner disk regions \cite{Kostaros:2024vbn,Astro_paper}. This avoids any  $''$contamination$''$ of strong lensing general relativity (GR) physics by accretion disk physics, which plagues the SMBH illumination by the continuum of the hot inner accretion disk. An optimal source for this can be the so-called broad line region (BLR) in active galactic nuclei, and more precisely its underlying cooler and mostly neutral phase \cite{Astro_paper}, hereafter designated as BLR$^{0}$ (as opposed to its ionized outer layers designated as BLR$^{+}$). With its unique spectral lines at IR wavelengths, and disklike gas distribution \cite{Thi:2024dny}, the BLR$^{0}$ can provide the photons to the light ring from $r=(10^2-10^3)R_s$ ($R_s=2GM/c^2$) away from the SMBH, without any such spectral line emission expected from anywhere nearer and toward the ISCO itself ($\lesssim 3R_s$). This unique far-field spectral line illumination of the SMBH keeps the background near the light-ring {\it ``emission-free''}. Imaging the  SMBH vicinity can then yield a $''$cleaner$''$ view of strong GR lensing effects. Furthermore, the dynamical timescales of the accretion disk scale with the mass of the SMBH and the distance from it. Thus, having emitting disk regions located further away from the SMBH naturally yields a less variable radiation background. The list of characteristics that position the  BLR$^{0}$ as a potent SMBH spectral line illuminator and  the astrochemistry responsible for its unique thermal state and its spectral lines at IR and near-IR wavelengths are the subject of dedicated papers \cite{Thi:2024dny, Astro_paper}, while an expanded discussion on the strong-gravity tests that can be conducted is laid out in \cite{PRD_companion}. In this work we focus our attention on two prominent effects unrelated to the light ring and related strong-gravity phenomena, which may be though the most accessible to future IR interferometers, i.e., the weaker lensing effect of the SMBH to the image of the BLR$^{0}$ and its spectrum. 

\emph{Illuminating a BH with the outer gas disk  ---} 
\begin{figure}
\begin{center}
\includegraphics[width=0.22\textwidth]{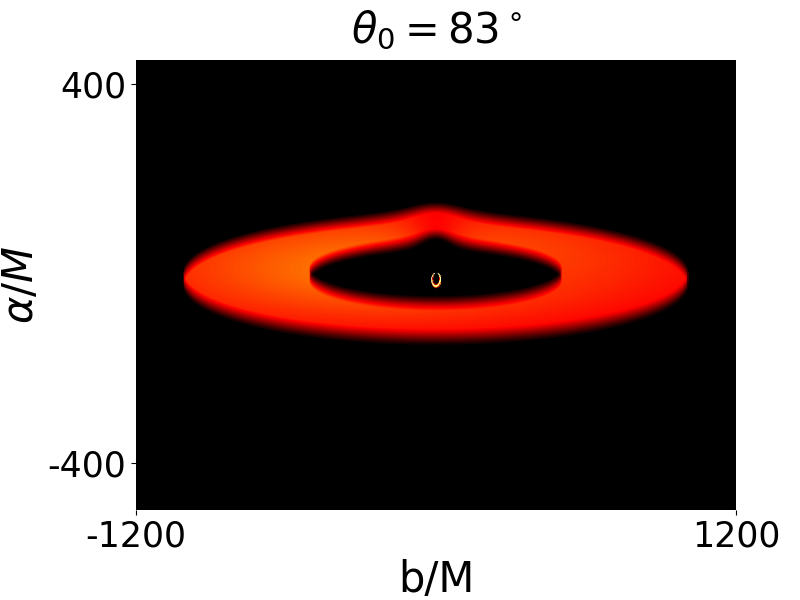}
\includegraphics[width=0.22\textwidth]{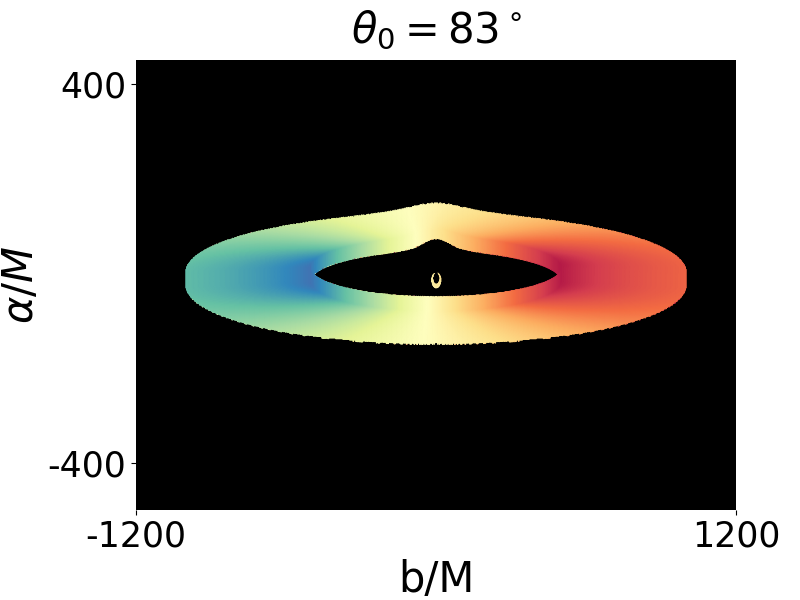}
\includegraphics[width=0.22\textwidth]{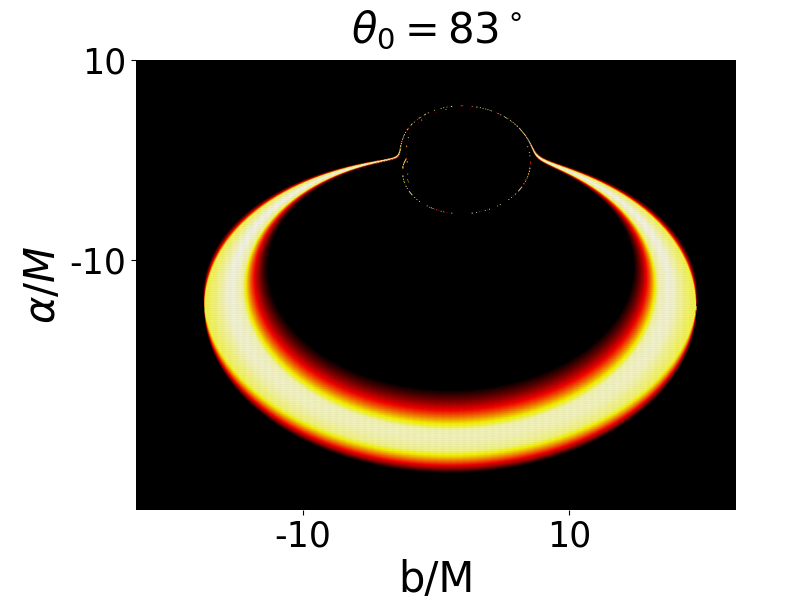}
\includegraphics[width=0.22\textwidth]{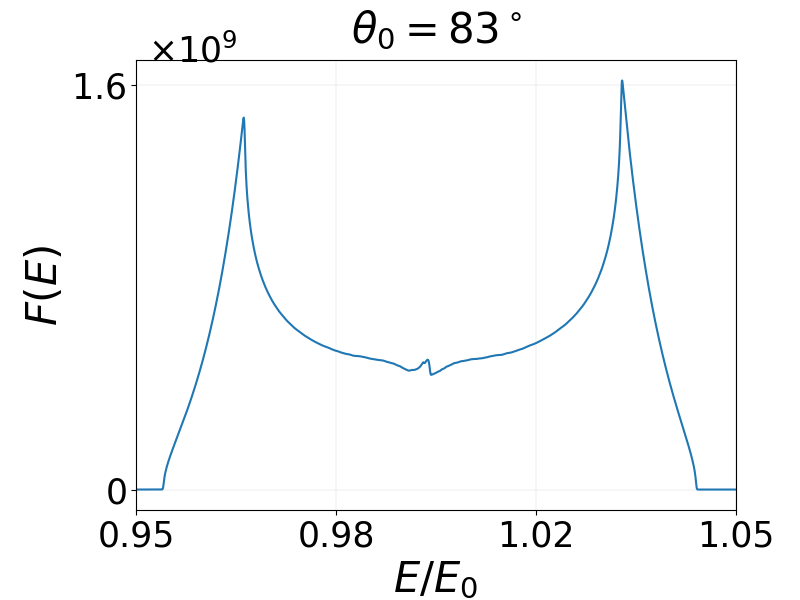}

\includegraphics[width=0.22\textwidth]{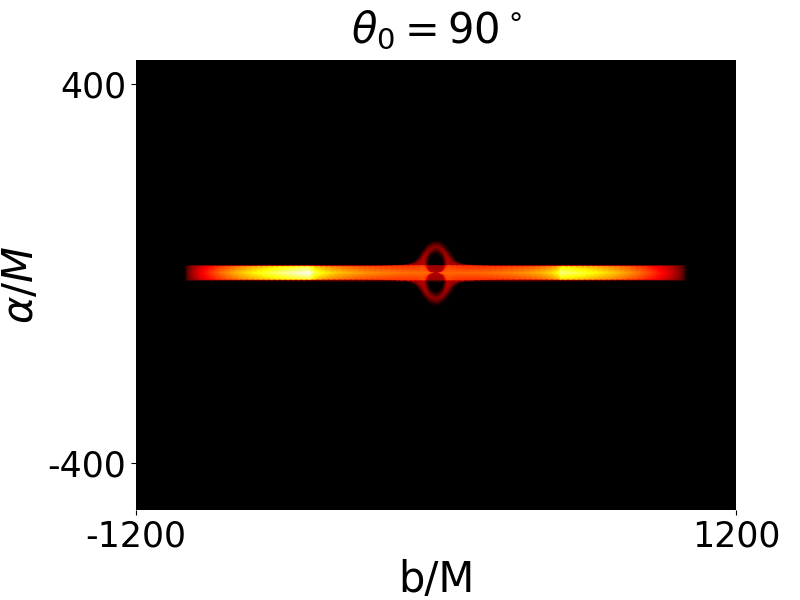}
\includegraphics[width=0.22\textwidth]{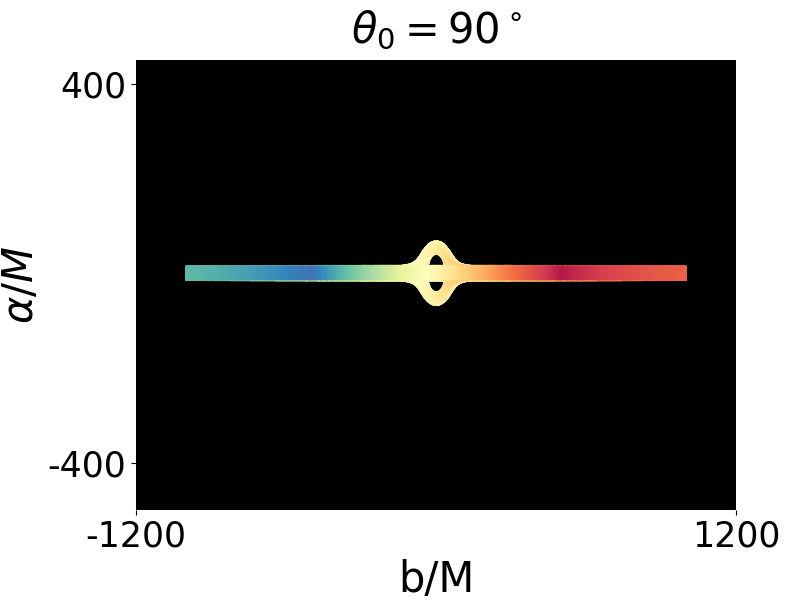}
\includegraphics[width=0.22\textwidth]{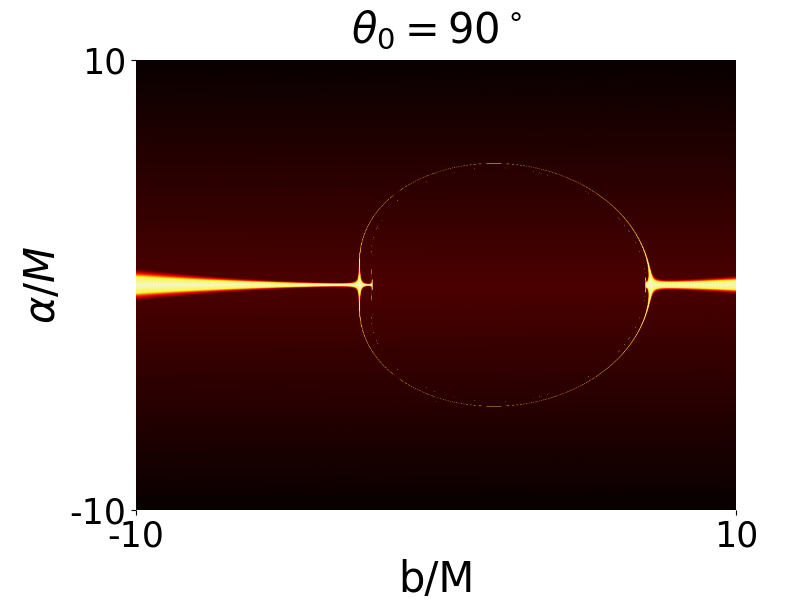}
\includegraphics[width=0.22\textwidth]{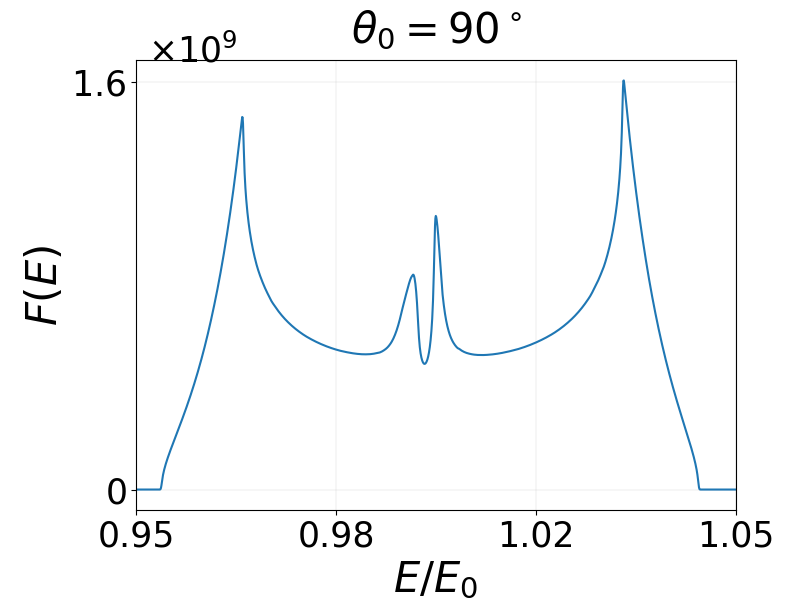}

\includegraphics[width=0.22\textwidth]{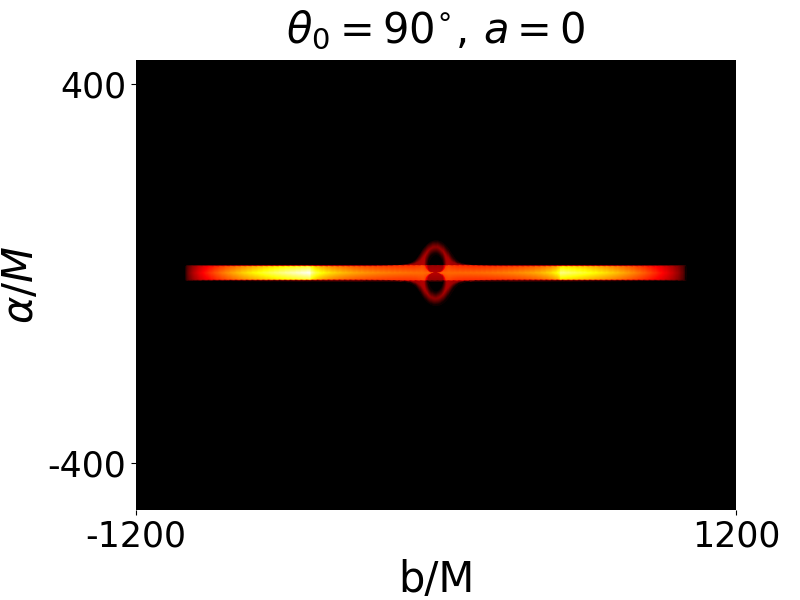}
\includegraphics[width=0.22\textwidth]{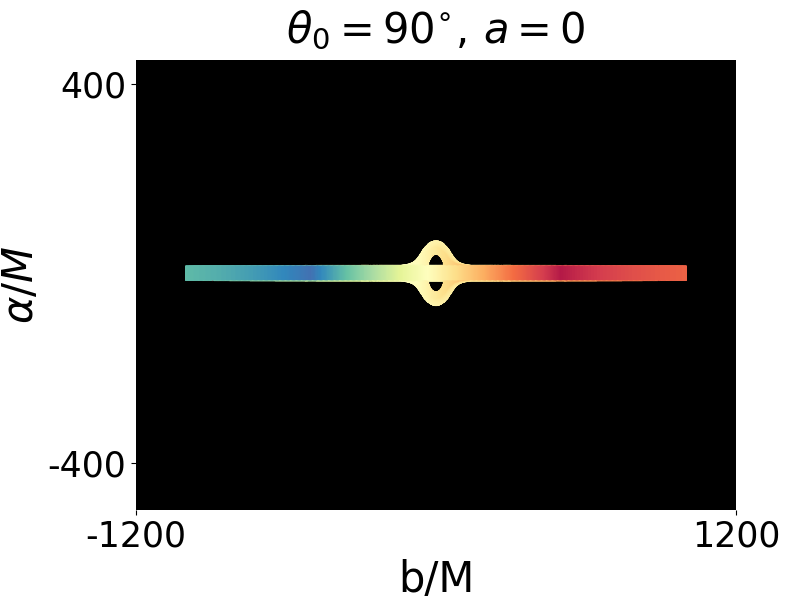}
\includegraphics[width=0.22\textwidth]{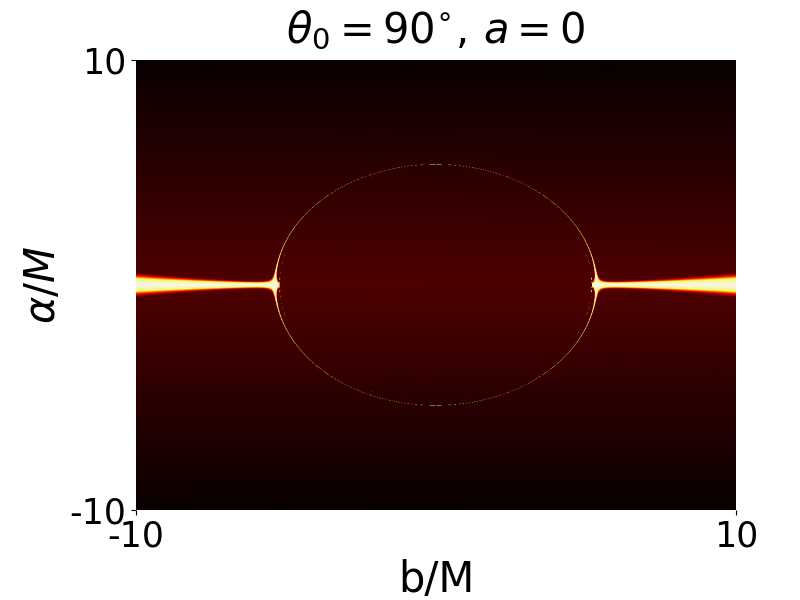}
\includegraphics[width=0.22\textwidth]{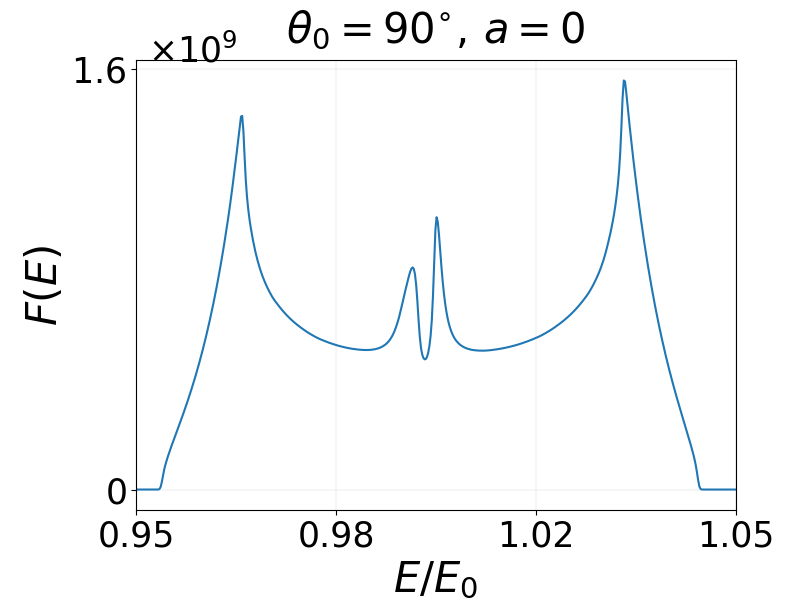}
\caption{Integrated intensity images, redshift maps, and observed spectra of an optically thin disk located between $500M<\varpi<1000M$ and $-15M<z<15M$. The first two rows correspond to Kerr BHs with spins $a=0.998M$ for viewing angles $\theta_0=83^{\circ}$, and $90^{\circ}$ (top to bottom), while the last row corresponds to a Schwarzschild BH at an angle of $90^{\circ}$.}
\label{fig:TT_bigscale}
\end{center}
\end{figure}
We demonstrate our idea by considering the illumination of a BH by a disk located at a radial distance between $r=500M$ and $r=1000M$ from the central object (using geometric units for the mass). In cylindrical coordinates ($\varpi,\varphi,z)$, the disk extends between $500M<\varpi<1000M$ and $-15M<z<15M$, so as to have a radius-hight relation of $H(\varpi)/\varpi<10\%$ \cite{Astro_paper}. We also assume a density profile for the emitting gas, which doubles as our local rest-frame emissivity $j_0$, that follows,
$j_0(r,\theta)=j_0\exp\left(-\frac{r-500M}{r_{\text{scale}}}\right)\exp\left(-\kappa\lvert \cos\theta\rvert\right)
$, with $r_{\text{scale}}=2000M$, $\kappa=10$,
and a velocity profile for the gas that follows,
$v^{\phi}=\left(\frac{\sqrt{M}}{(r\sin\theta)^{3/2}+a\sqrt{M}}\right)$,
which describes an almost Keplerian disk ($a=J/M$ is the Kerr parameter and $(r,\theta)$ the spherical coordinates). These are typical choices in the literature for the density profile (with a simple power law being the most common) and  will suffice for our demonstration \cite{Younsi2012,Fuerst_2004,Fuerst_2007,Storchi_Bergmann_2017}. Furthermore, we assume the disk emission to be spectral line emission at the rest frame (of the fluid) frequency $\nu_0$. The line frequency is kept arbitrary and the results are expressed in terms of $\nu/\nu_0$ (with $\nu$ the observed frequency). The final assumption is that the disk is {\it effectively} optically thin, with a turbulence-induced velocity dispersion of $\Delta V_{\rm turb}=150 km/s$ \cite{Astro_paper}. Thus, the line profile in the gas rest frame is not a delta function (as frequently assumed in past studies) but a Gaussian centered at $E_0$ with $\sigma=5\times10^{-4}E_0$.

In order to produce the expected line brightness image, and emergent line profile, we integrate the radiative transfer equations of the Lorentz Invariant intensity $\mathcal{I}_\nu=I_\nu/\nu^3$, given by \cite{Younsi2012,Fuerst_2004,Fuerst_2007},
\begin{equation}
    \frac{d\mathcal{I}_\nu}{d\lambda}=\gamma^{-1}\left(\frac{j_0}{\nu^3_{0}}\right),
\end{equation}
where $\gamma=\nu/\nu_0$, with $"$0$"$ indicating quantities in the local rest frame of the emitting gas. We assume zero absorption along the geodesic, which is realistic for the macroturbulent BLR velocity fields  \cite{Astro_paper}. We calculate along any geodesic the specific intensity in terms of the observed frequency $\nu_{obs}$,
    $dI_\nu=d\mathcal{I}_\nu \nu_{obs}^3=\gamma^{-1}\left(\frac{j_0}{\nu_0^3}\right)\nu^3_{obs}d\lambda$,
which gives the contribution to the specific intensity from every step along each geodesic \cite{Fuerst_2004}, provided by ray-tracing \cite{Kostaros_2022,Kostaros:2024vbn}. 
Here we use the specific intensity in terms of photon energy $I_E$ instead of $I_\nu$. The specific flux is then given by integrating the solid angle over which the source is viewed, 
    $dF_E = I_Ed\Omega.$
For our configuration, the solid angle on the observer's screen is 
    $d\Omega=db d\alpha/D_L^2$,
where $D_L$ is the distance from the observer to the BH (flat-space distance), while $b$ and $\alpha$ are the impact parameters on the observer's screen with respect to the BH (in units of the BH mass) \cite{Kostaros:2024vbn}. 

For the BH, we use either Kerr BHs rotating at the Thorne limit \cite{1974ApJThorne} with $a=0.998M$ and $M=1$ (the mass is our length scale) or, for comparison, Schwarzschild BHs. The disk is on the equatorial plane, whereas the observer is placed at various angles with respect to the axis of rotation of the BH.

\emph{Disk, shadow, and spectra  ---} 
%
%
In \cref{fig:TT_bigscale} (from left to right) we show the integrated (bolometric) intensity image of the torus, the redshift map, a zoom into the strongly lensed image of the torus coming from the central region of the BH (near the light ring), and the total observed line profile produced by both the direct emission from the disk and the strongly lensed emission that the BH redirects towards the observer. These are shown for viewing angles of $\theta_0= 83^{\circ}$, and $90^{\circ}$ for the top two rows. The last row shows a Schwarzschild BH at $90^{\circ}$. 

There are a few interesting things to note in these figures. In the intensity images, the lensing effect the BH produces on the image of the far side of the torus is clear. In fact, what we observe is the formation of an hourglass-like feature, which we dub as Einstein hourglass \cite{footnote1}, around the BH, due to the emission from the far side of the disk that is deflected in an Einstein-ring-like pattern. 
There are some interesting features to observe in the spectra as well, which are more evident at edge-on orientations. The photons coming directly from the disk are affected by the kinematics of the gas, depending on the projected orbital velocity along the line of sight, and therefore present Doppler and beaming effects. This yields a spectrum that is broadened, like the one we see on the top row of \cref{fig:TT_bigscale} that corresponds to $\theta_0=83^{\circ}$, where we observe the usual double-peak line profile \cite{Storchi_Bergmann_2017}.

\begin{figure}[h]
\begin{center}  
\includegraphics[width=0.3\textwidth]{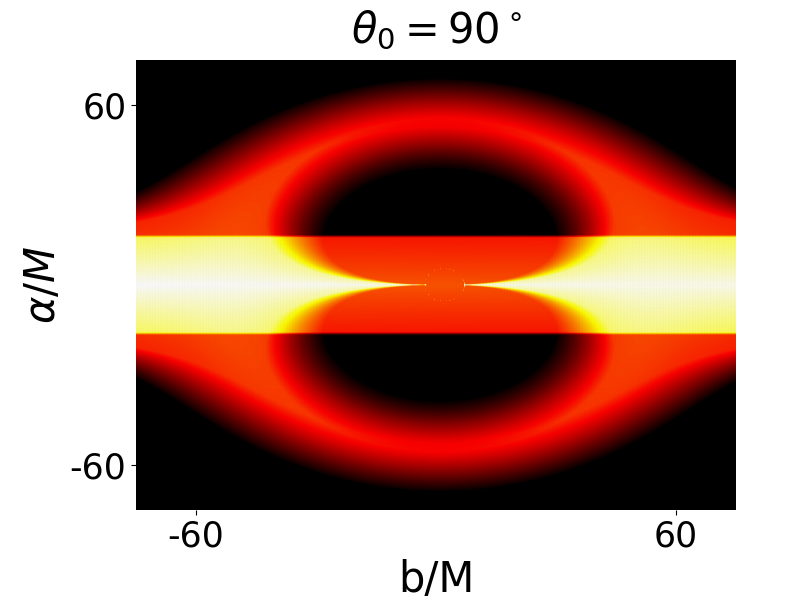} 
\includegraphics[width=0.3\textwidth]{hist_theta90.png} 
 \caption{Close-up of the {\it Einstein Hourglass} feature and the spectrum of the edge-on Kerr BH accretion disk of \cref{fig:TT_bigscale}.}
 \label{fig:ring1}
\end{center}
\end{figure}

For edge-on orientations an additional feature shows up in the global spectrum: {\it a middle double peak due to the Einstein hourglass lensing}. The two peaks are near the rest-frame frequency, suffering only some minor gravitational redshift due to the BH gravitational potential at the location of the disk and some Doppler effects (the cause of the split into two peaks) related to the formation mechanism of the Einstein hourglass. This line brightness feature and its spectrum are shown in more detail in \cref{fig:ring1}. Their prominence depends on the inclination and becomes more so for edge-on orientations, as expected from the lensing geometry.
This unique spectrum feature appearing in the total line profile could provide a smoking-gun signal of lensing by the SMBH in galactic centers and of the existence of an underlying Einstein hourglass, even if the later remains unresolved by the particular observations. 

\emph{Einstein hourglass and its spectral counterpart ---} 
%
%
\begin{figure}[h]
\begin{center}
    \includegraphics[width=0.6\textwidth]{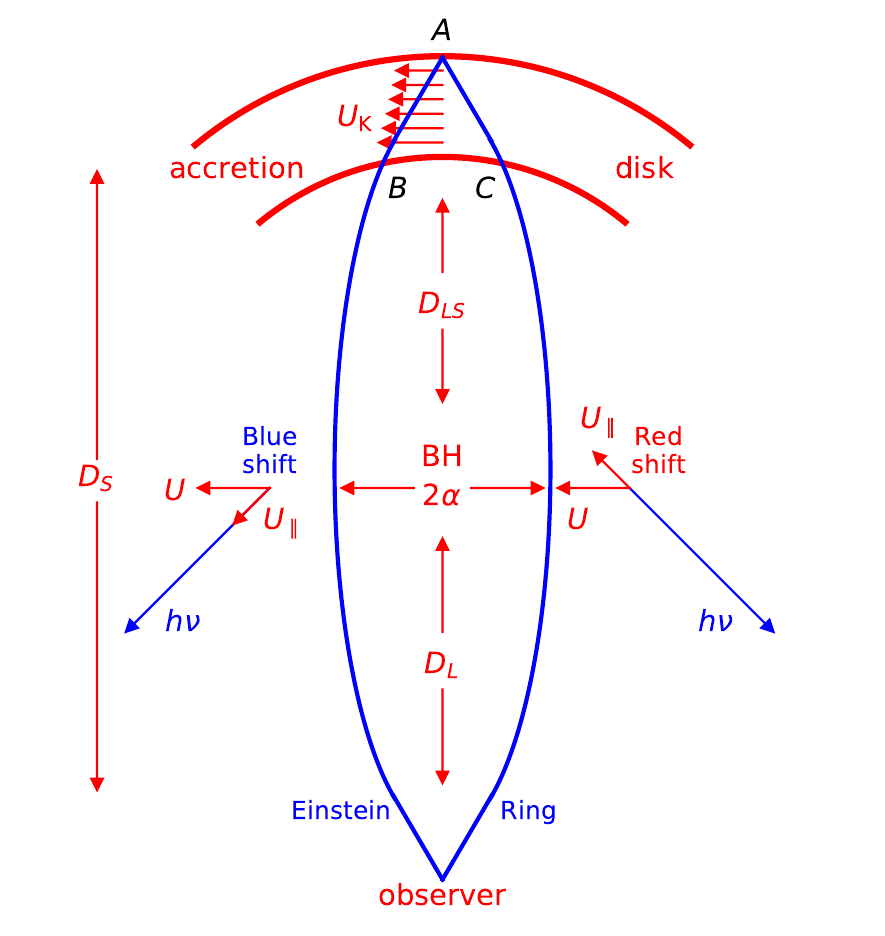}
 \caption{Cartoon of the geometry of the emission coming from behind the BH, that forms an Einstein ring.}
 \label{fig:cartoon1}
\end{center}
\end{figure}
%
\begin{figure}
\begin{center}   \includegraphics[width=0.6\textwidth]{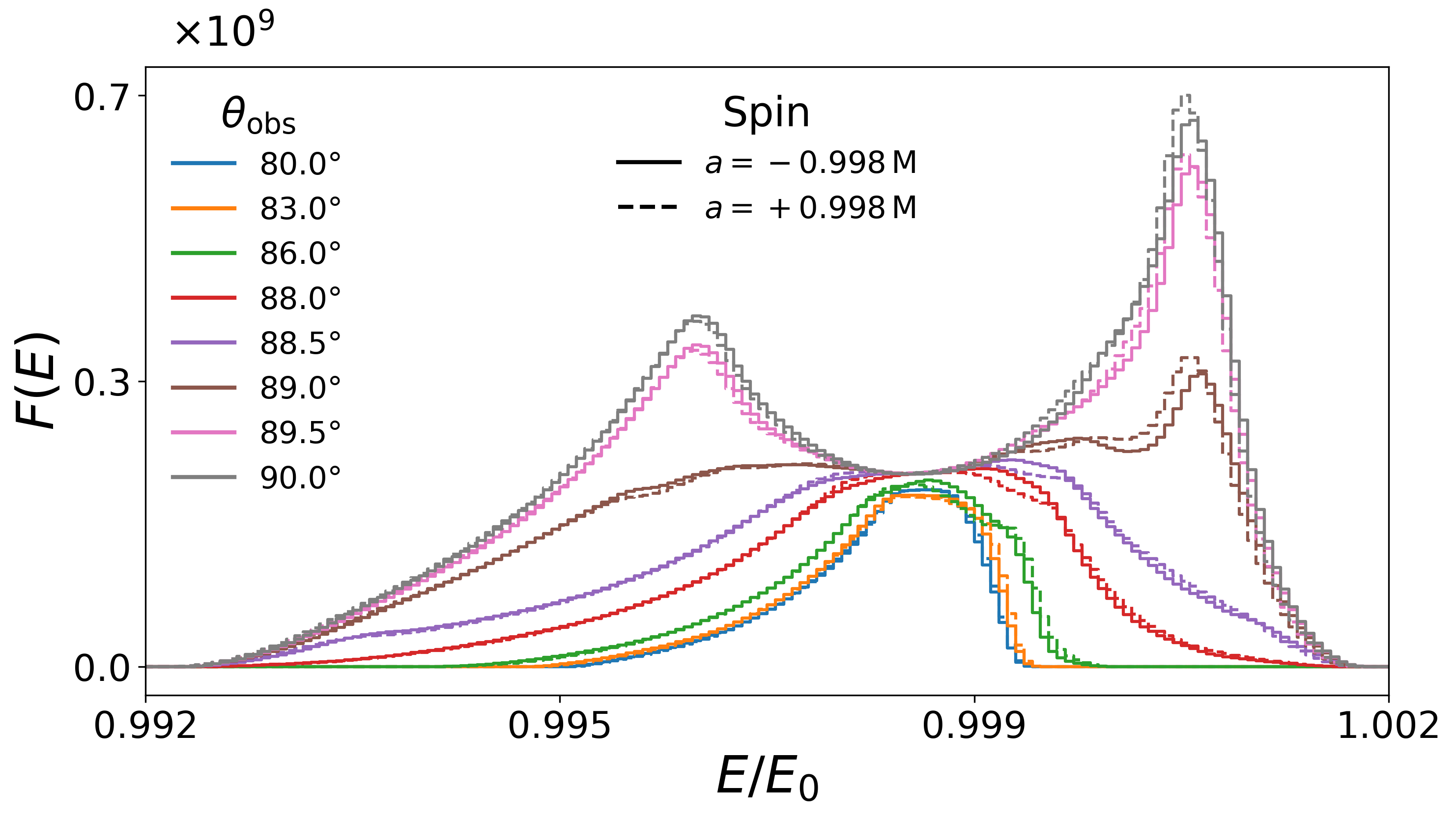}
 \caption{Demonstration of the emergence of the central magnified double peak of the spectral line, by isolating the emitting region behind the BH as shown in Figure \ref{fig:cartoon1}, for two BH spins.}
 \label{fig:E_spectra}
\end{center}
\end{figure} 
%
We will now give a brief analytical demonstration how the Einstein hourglass feature yields the central feature seen in the global line profile. 
We assume the central SMBH as a point mass (given the size of the disk, this is a reasonable assumption). Then, when the disk is viewed in an edge-on configuration, the light rays emanating from the emitting material directly behind the BH would form an Einstein ring (see \cref{fig:ring1}). The ring's angular size, for an illumination source at distance $D_S$ and a lens at distance $D_L$ from an observer, and a geometry like the one given in \cref{fig:cartoon1}, is given by the expression \cite{2018bookCongdon},
\be 
\theta_E=\sqrt{\frac{D_{LS}}{D_L D_S}\frac{4GM}{c^2}},
\ee
where $D_{LS}$ is the source-lens distance (assumed to practically be in flat space). For our given geometry, where $D_S\simeq D_L\gg D_{LS}$, this would give 
\be \label{eq:alpha}
\alpha_E= D_L \theta_E=2\sqrt{D_{LS} G M/c^2},
\ee
where  $\alpha_E$ is essentially the impact parameter of the photons and the radius of the Einstein ring on the observer's image plane.
For  a disk radius  $500M<\varpi<1000M$, it will be: $45M<\alpha_E<63M$, as seen in \cref{fig:ring1}. In this configuration, apart from the formation of the band of Einstein rings, the emission from behind the BH and along the line of sight is also magnified due to the lensing effects, just as a point-mass lens magnifies a background star \cite{2018bookCongdon} (with the $''$star$''$ now being the far side of the disk), forming in this manner the central peaks on the total line profile~(\cref{fig:ring1}) and the hourglass effect \cite{PRD_companion}. 

To verify this picture, we calculated the spectrum from a block of emitting matter, using ray-tracing, with a square cross-section that extends from $\varpi_{in}$ all the way to $\varpi_{out}$, located behind the BH. 
The resulting spectra in \cref{fig:E_spectra}, show the emergence of two peaks as we move to edge-on inclinations, where the emission is magnified and also shifted from the actual emission frequency $E_0$. What is causing the frequency split of the inner double-peak? As \cref{fig:cartoon1} shows, the rays go through the gas disk at some angle. 
Thus, two peaks, one redshifted and the other blueshifted. are obtained because every ray intersecting the line of sight along the back side of the disk that contributes to the formation of the Einstein hourglass will move through the gas either in the same sense as it moves around the BH (left ray in \cref{fig:cartoon1})or in the opposite sense (right ray in \cref{fig:cartoon1}). 
%

The rays following a polar orbit will be the only ones without any energy shift, but these are only a small fraction of all the light rays, providing less flux in the middle between the two peaks. One can estimate the relevant Doppler shifts, assuming a thin lens approximation, where the impact parameter of a light ray coming from behind the BH is equal to the distance of the observer from the BH times the angular size of the Einstein ring, which will be $\theta_E D_L$. Since in this approximation, the light rays are straight lines, the ray intersects the gas element of the disk (moving horizontally) at the point of emission, at an angle that will result in a projected velocity along the ray equal to 
\be u_\parallel = u_K \frac{\theta_E D_L}{D_{LS}}=\sqrt{\frac{GM}{D_{LS}}}\sqrt{\frac{4GM}{c^2D_{LS}}}=\frac{2GM}{c D_{LS}},
    \ee
which will be either in the direction of the emission (for the rays on the left) or against the direction of emission (for the rays on the right). The total Doppler shift between left and right rays will then be, 
\be \label{eq:dnu}
     \frac{\Delta \nu}{\nu} = 2 \frac{u_\parallel}{c} =\frac{4GM}{c^2 D_{LS}}, 
    \ee
which for radii between $500M<r<1000M$  gives shifts between $\frac{4}{1000}<\frac{\Delta \nu}{\nu}<\frac{4}{500}$. This can be seen in  \cref{fig:ring1} (and \cref{fig:E_spectra}) where the Doppler shift from the emission coming from the far side of the accretion disk (at $D_{LS}=r_{out}$) corresponds to the energy shift between the two central peaks, while the Doppler shift from the emission coming from the near side of the accretion disk (at $D_{LS}=r_{in}$) corresponds to the total width of the central magnified part of the line. 
This line splitting, due to the presence of the lens, is essentially measuring the velocity of the source perpendicular to the line of sight. 
The resulting spectra have little dependence on the BH spin, and primarily depend on the inclination, which determines the eventual lensing magnification  (see \cref{fig:E_spectra}).

\emph{Conclusions  ---} 
%
In this work we considered spectral line radiation from a BLR$^{0}$ gas disk illuminating the strong-gravity region of a BH, following Refs. \cite{Kostaros:2024vbn,Thi:2024dny,Astro_paper}. Such an illumination source yields several advantages in probing strong-gravity effects near a BH, detailed in full in \cite{PRD_companion}. Here we examined the two most prominent ones, and more readily accessible to future IR line imaging observations. We conclude by showing that a combination of these two effects can yield SMBH mass estimates. 

If the hourglass feature size $\alpha_E$ (corresponding to the size of the Einstein ring) is measured \eqref{eq:alpha} (which will be easier to do than the $\sim$10 times smaller light ring), and from spectroscopic measurements the $\Delta \nu/\nu$ of the double central peak is obtained \eqref{eq:dnu}, they can be combined to yield the BH mass $M$ from 
\be \frac{GM}{c^2}=\left(\frac{\alpha_E^2}{16} \frac{\Delta \nu}{\nu}\right)^{1/2},\ee 
(and the  $D_{LS}$ distance of the emitting disk). This can be a novel way of measuring SMBH masses, and a beautiful, new demonstration of lensing in galactic centers. The strong-gravity effects associated with the much smaller light ring that can be probed using such a far-field illuminator, and their observability, will be presented in following work \cite{PRD_companion, Astro_paper}. Finally we note that at the operating frequencies of the EHT ($\sim $230\,GHz and 350\,GHz in its upgrade), the corresponding BLR$^{0}$ thermally excited spectral lines will be undetectable \cite{Astro_paper}, and thus EHT high-resolution imaging will likely miss these effects.

\emph{Acknowledgements---} 
%
%
This work has received no funding. 
PP would like to thank the staff at the European Southern Observatory (ESO) Headquarters in Munich for their hospitality, during which some of these ideas were developed, along with the bartenders of a certain bar called Schumman's at Odeonplatz. 

\emph{Data Availability---} 
The data that support the findings of this study were generated by numerical simulations. The ray-tracing and radiative transfer code used to produce the images and the spectra is publicly available \cite{dataGIT}.

\bibliography{bibliography.bib}

\begin{thebibliography}{45}
\bibitem [{foo()}]{footnote1}%
  \BibitemOpen
  \href@noop {} {}\bibinfo {note} {To name it an Einstein ring would be a
  misnomer, as the horizontally extended illumination source, unlike a
  star-like point source, creates something more complicated than a ring
  \cite{PRD_companion}.}\BibitemShut {Stop}%
\end{thebibliography}

\end{document}